# A Novel Advanced Heap Corruption and Security Method

Arundhati Walia[1], Prof (Dr.) Syed i. Ahson[2]

1Computer Science Department, Shobit University

Meerut, UP, India.

sameer_walia00001@yahoo.com

2 Computer Science Department, Shobit University

Meerut, UP, India.

drsiahson@yahoo.com

**Abstract**

Heap security has been a major concern since the past two decades. Recently many methods have been proposed to secure heap i.e. to avoid heap overrun and attacks. The paper describes a method suggested to secure heap at the operating system level. Major emphasis is given to Solaris operating system's dynamic memory manager. When memory is required dynamically during runtime, the SysVmalloc acts as a memory allocator.Vmalloc allocates the chunks of memory in the form of splay tree structure. A self adjusting binary tree structure is reviewed in the paper, moreover major security issue to secure heap area is also suggested in the paper.

## 1. Introduction

Trustworthiness in the computing has mainly focused on secure software which is free from every possible cause of attacks due to vulnerabilities that exist in the software. Therefore major approach of trustworthy computing is to deliver secure and reliable computing experience [1]. Buffer overflow attacks have been a major concern for software security wherein a malicious user of software feeds data to the fixed length buffer that exceeds beyond the size of buffer. Hence an attacker can get unauthorized access to the computer causing memory exploits. Moreover buffer overflow attacks are major cause of security issue in the modern operating systems too. The various categories of buffer overflow attacks are:

    (i)     Stack overflow.
    (ii)    Heap overflow
    (iii)   Integer overflow etc.

Major concern is given to Heap overrun and how heap area created by the programmer can be prevented from getting overrun. Section 1 of the paper describes an example of heap overrun. Section 2 describes the Vmalloc's function for memory allocation and self adjusting tree structure. Section 3 explains various operations performed on splay tree as well as heap security method and section 4 concludes the paper.

**Heap overflow**

Heap is a region of virtual memory used by the program. The memory manager allocates heap memory area in the form of chunks.dlmalloc and SysVmalloc are popular heap applications when an application doesn't know the size of the object in advance. Heap space can be created by programmer using functions like malloc () and new (). Many systems are written in the languages like C/C++ and even most of the projects are implemented in unsafe languages like C/C++ which provides no boundary checks for the functions like strcpy(),strcat() etc.For instance strcpy(char *dstn,char *src) copies string pointed out by source to destination, It is only left upon the programmer to check whether the destination buffer is sufficiently large to accommodate source string otherwise this may cause memory corruption altering the program flow control thus breaking system security. If the destination buffer is allocated on the heap then heap overruns causing heap overflow attack. Simple example of heap overflow [2].

              a=malloc (96);
              b=malloc (80);
              c=malloc (80);
            strcpy (a, argv [1]);
                free (a);
                free (b);
                free (c);

## 2. Dynamic memory allocater for Solaris operating system

On modern environments it is increasingly desirable to use shared memory to speed up the process communication and mapped memory for faster I/O persistence.malloc is not designed for this purpose.Vmalloc enables application to allocate arbitrary type of memory and pick allocation strategies that matches allocation requirements. Allocation of the memory is performed by the following calls [3].

(i) Vmalloc (Vm, size): Allocates from the region Vm a block of size bytes.
(ii) Vmfree (Vm, b): Makes the previously allocated space for block b available for future allocation.
(iii) Vmresize (Vm, b, size_type): resizes block b to size bytes.
(iv) Vmclear (Vm): is useful to globally free all currently busy blocks in the region.

Vmbest is general purpose allocator. The basic allocation strategy is best fit i.e. any allocation request is satisfied from smallest free area. Free areas are kept in splay tree.

### 2.1 Tree structure

Larger chunks both free and allocated are arranged in tree like structure. Each node contains list of the chunks of the same size [4].
Declaration: Structure of node in a free tree

```
typedef struct_t_{
    word t_s        /*size of this element*/
    word t_p        /*parent node*/
    word t_l        /*left child */
    word t_r        /*right child */
    word t_n        /*next in link list*/
    word t_d        /*dummy to reserve space for self pointer*/
} TREE;
```

Splay tree: Splay tree structure is shown in figure 1.

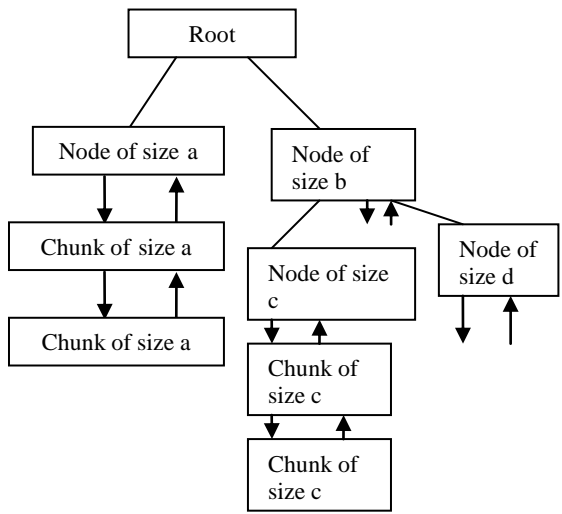

Fig.1 Splay Tree

When Vmalloc allocates a memory, following chunks of memory is created as shown in figure 2.

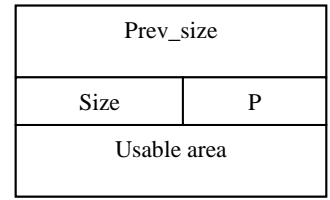

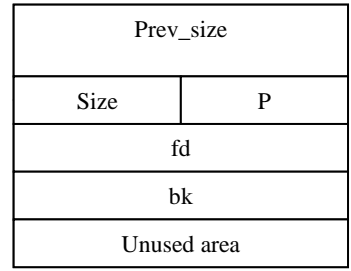

Fig.2 allocated and free memory chunks

When dynamic memory manager manages the heap area, major cause of vulnerability is due to data pointers like [5].
i) One of the data pointers can be overwritten with the value in another data pointer.
ii) Two pointers used for linking free chunks.
iii) Freeing the chunk twice.

We considered address protection method in software where address encoding method is suggested. SysVmalloc uses self adjusting binary tree for storing the information. Each node contains the list of chunks of same size. The various chunk sizes are shown in fig 3.

Fig.3 Chunk sizes

| structure | Chunk size | Node size |
|---|---|---|
| Singly link list | Uniform | Size a |
| Doubly link list | Uniform | Size b |
| Doubly link list | Variable(sorted) | Size c |
| Doubly link list | Variable(unsorted) | Size d |

### 2.2. Chunk sizes of node of a tree

Figure 4 shows self adjusting binary tree structure showing various chunk sizes of each node of tree.

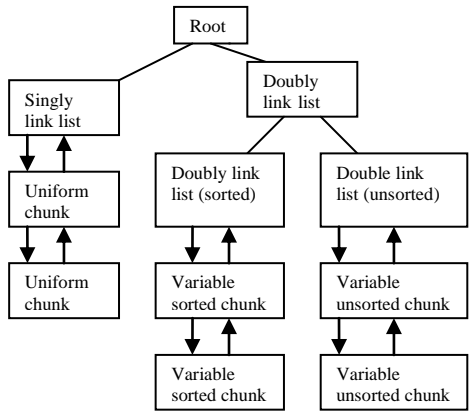

Fig .4 chunk sizes of each node of a tree.

Self adjusting data structure has following advantages [8].
1. They are more efficient since they adjust according to the usage.
2. They need less space.
3. Their access and update algorithm are conceptually simple and easy to implement. The taxonomy of various buffer overflow attacks [6] is shown in the figure 5.

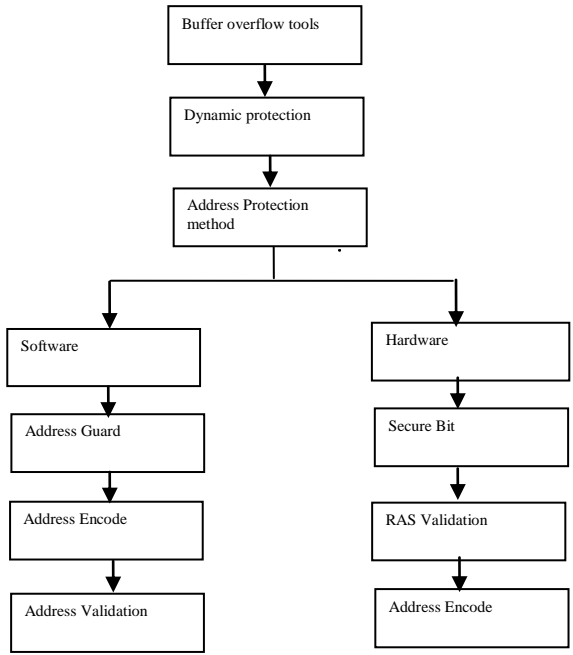

Fig. 5 Taxonomy of solutions against overflow attacks

## 3. Operations performed on Splay tress and proposed Encoding Method

The chunks of various sizes as a nodes of the tree are shown in figure 4.In self adjusting binary tree, the node which is of interest in brought towards the top i.e. root by using certain rotations from node to root because of which the nodes which are of least use are kept in the bottom of the tree. The only elements that are utilized in the nodes of the tree are $t\_s$, $t\_p$ and $t\_l$ elements [7]. When the chunk is allocated, Vmalloc tries to take a free chunk from the tree. The various operations performed on the splay tree are [9].
a) insert (chunk P, t): Insert a chunk P in a tree t.
b) delete (chunk P, t): delete chunk P from tree t.Insertion and deletion can be performed using join and split.
c) join (t1, t2): join is performed by accessing largest item i in t1 and thus has NULL right child, the join operation is completed by making t2 right sub tree and returning resultant tree.
Let t1 and t2 be two trees:

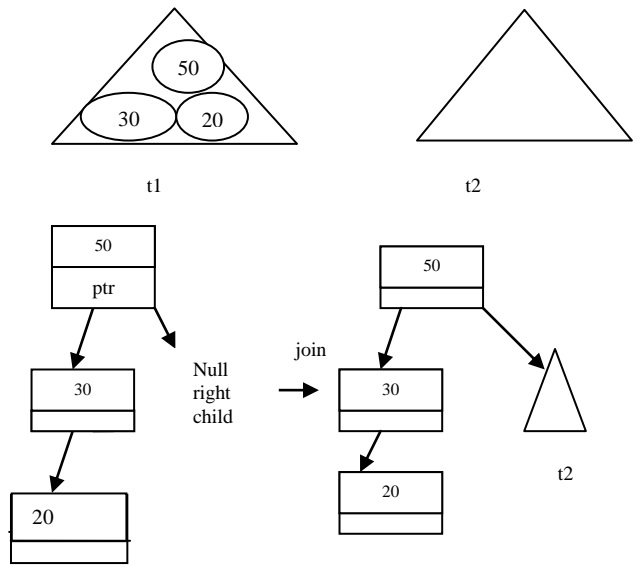

Fig .6 Join operation

d) split (i, t): perform access (i, t) and then return two trees formed by breaking either left link or right link from the new root of tree.

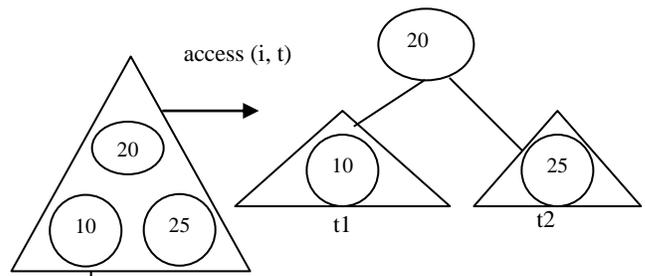

Fig. 7 Access operation

insert(i,t):perform split(i,t) and then replace t by tree consisting of new root node having i whose left and right sub trees are trees t1 and t2. To perform insertion of node 28.

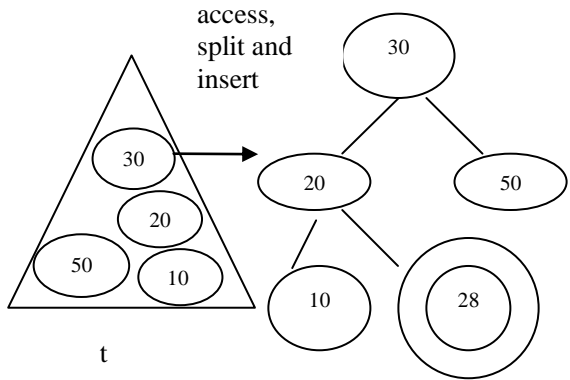

Fig 8. Insertion

delete (i, t): To perform delete of i, access item i and replace t by join of its left and right sub trees.

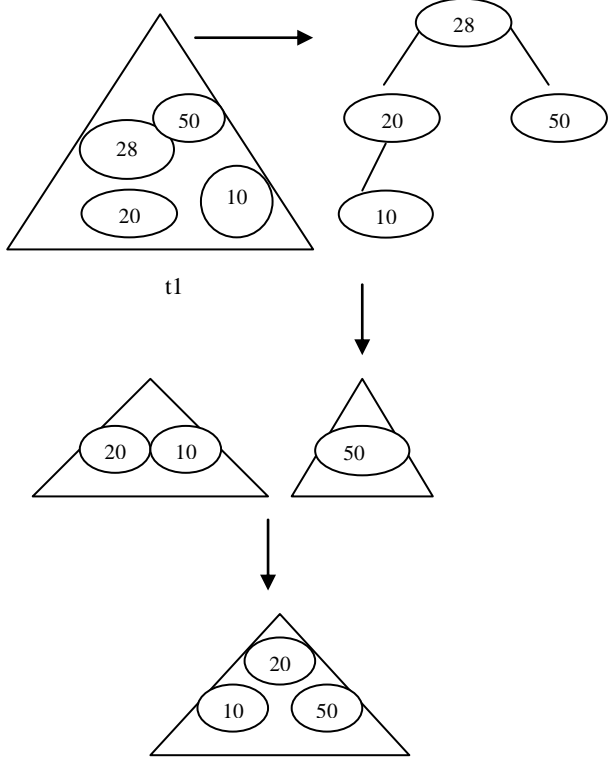

Fig .9 Deletion operation

When chunks are freed they are added to the special array that holds upto 32 chunks. If one of the chunk is previously overflowed, so that it now contains crafted fake chunk provided by an attacker, the process of consolidating it would lead to arbitrary memory overwrite causing overflow attack in heap area during dynamic allocation. To avoid such types of overflow, the data pointers pointing to the chunks should be protected. The pointers are secured by encoding them and then decoding them back when required data. There are several cryptographic key generating algorithms which help to encode the pointers. The advance Cryptographic algorithm like RSA [10] can be used to encode the address. Tree structures generally speed up the process. The encoded chunk of tree structure is shown in figure 10. According to RSA Algorithm, the pointers pointing to the next chunk is encoded by converting the address into the number.

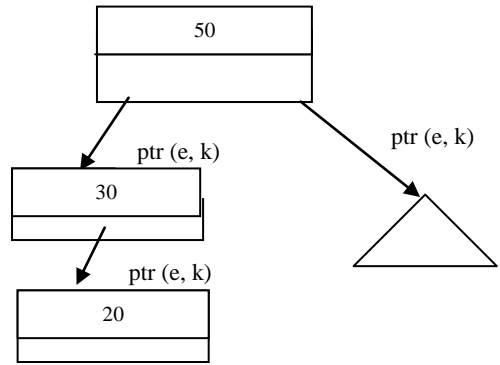

Fig .10 Nodes with encoded pointers

## 4. Conclusion

The paper summarizes the major area of software to have trustworthiness. When SysVmalloc, a dynamic memory manager allocates the heap area in the forms of memory chunks represented by self adjusting binary tree structure, the free and allocated chunks of memory may get overwrite causing overflow attacks. The taxonomy of various methods for prevention is shown. Moreover the paper covers the means suggested to encode the data pointers pointing to the free and allocated chunks. The further scope of the paper is to use cryptographic algorithm to perform encoding.